\documentclass[12pt,english]{article}
\usepackage[T1]{fontenc}
\usepackage[latin9]{inputenc}
\usepackage{geometry}
\usepackage{units}
\usepackage{amstext}
\usepackage{amssymb}
\usepackage{graphicx}
\usepackage{setspace}
\makeatletter

\newcommand{\noun}[1]{\textsc{#1}}
\providecommand{\tabularnewline}{\\}
\newcommand{\lyxdot}{.}

\newcommand{\lyxaddress}[1]{
	\par {\raggedright #1
	\vspace{1.4em}
	\noindent\par}
}

\makeatother

\usepackage{babel}
\begin{document}

\title{Noise in gene expression may be a choice of cellular system}

\author{Rajesh Karmakar\thanks{Electronic address: rkarmakar@vidyamandira.ac.in}}
\maketitle

\lyxaddress{\begin{center}
Ramakrishna Mission Vidyamandira, Belur Math, Howrah -711202, West
Bengal, India
\par\end{center}}
\begin{abstract}
Gene expression and its regulation is a nonequilibrium stochastic
process. Different molecules are involved in several biochemical steps
in this process with low copies. It is observed that the stochasticity
in biochemical processes is mainly due to the low copy number of the
molecules present in the system. Several studies also show that the
nonequilibrium biochemical processes require energy cost. But cellular
system has developed itself through natural evolution by minimizing
energy cost for optimum output. Here we study the role of stochasticity
qualitatively in a network of two genes using stochastic simulation
method and approximately measure the energy consumption for the gene
expression process. We find that the noise in gene expression process
reduces the energy cost of protein synthesis. Therefore, we argued
that the stochasticity in gene expression may be a choice of cellular
system for protein synthesis with minimum energy cost.

\end{abstract}

\section{Introduction}

Gene expression (GE) is a basic cellular process whereby proteins
are synthesized according to the nucleotide sequences in the gene.
Gene expression involved several biochemical reactions, the kinetics
of which determine how the number of participating biomolecules changes
as a function of time. There are two major steps in gene expression,
transcription and translation. In the process of transcription mRNAs
are synthesized. During the process of translation, the sequence of
mRNA molecule is translated into the proteins \cite{key-1}. Regulation
is ubiquitous in complex living system. Gene expression is a regulatory
process which can takes place either at transcription, translation
or degradation levels. Many regulatory molecules are involved in the
gene expression process. There are two major types of regulatory molecules:
activator and repressor \cite{key-2}. Both types of regulatory molecules
are also proteins and synthesized from some other genes. The regulatory
molecules which regulate the transcription process are called the
transcription factors (TFs). They bind the specific binding site/sites
on the gene and drive the gene into a state called the ``ON'' state
or active state. The unbound state is called the ``OFF'' state or
inactive state. Under the regulation by TFs, the gene can be either
in the ON or OFF state depending on whether the TFs are bound to the
gene or not \cite{key-1,key-2,key-3,key-4}. If the gene is in the
ON (OFF) state then the transcription process can takes place and
mRNAs are synthesized with higher (low/basal) rate \cite{key-4}.
From the newly born transcripts/mRNAs, proteins are synthesized and
the process is called the translation. In the synthesis of proteins,
ribosomes play the important roles. There are specific binding sites
on the newly born mRNAs where ribosomes can bind and synthesize proteins.
Newly born proteins have a specific degradation rate and the RNAse
molecules do that job. Each and every biomolecules do not exist forever
to work rather they have a specific degradation rate. It is theoretically
and experimentally well established that the biochemical events in
gene expression are inherently stochastic in nature \cite{key-5,key-6,key-7,key-8,key-9,key-10,key-11,key-12,key-13,key-14,key-15,key-16,key-17,key-18,key-19}.
The timing of the biochemical events cannot be predicted with certainty.
The stochasticity in biochemical events lead to the fluctuation in
mRNA and protein levels about a mean value. That fluctuation is called
the noise in mRNA/protein level. Several studies show that the noise
in gene expression appears due to low copy number of molecules (e.g.,
small number of regulatory molecules, very low gene copy number etc.)
involved in the gene expression and regulation process \cite{key-11,key-12,key-13,key-16,key-17,key-18,key-19}.
This stochastic nature of the biochemical reactions may be ignored
in the limit of large numbers of biomolecules. The noise in gene expression
appears from the random switching between the ON and OFF states, random
production and degradation of mRNAs and proteins \cite{key-5,key-6,key-7,key-8,key-9,key-20,key-23}.
It has also been shown that the stochastic effects due to random transitions
between ON and OFF states of a gene are much stronger than the stochastic
effects caused by random production and degradation of single mRNA
and protein molecules \cite{key-9,key-10,key-14,key-15}. That happens
because low copy number of regulatory molecules are involved to regulate
one (haploid) or two (diploid) copies of a gene. 

Several studies show that the stochasticity in gene expression produces
two types of responses: graded and binary \cite{key-24,key-25,key-26,key-27,key-28,key-29,key-30}.
In graded response, the mean protein level changes gradually and the
distribution of protein level shown to be unimodal. In binary response,
gene expression occurs either at low level or at high level and the
distribution of proteins will be bimodal. The bimodal distribution
of protein level generally observed with positive feedback network
\cite{key-31,key-32}. But, stochasticity itself can produce bimodal
distribution without any positive feedback loop. Random switching
between the ON and OFF states of the gene play the important role
in generation of bimodal distribution of protein level when there
is no feedback loop \cite{key-26,key-27,key-28,key-29,key-30}. 

The presence of noise or fluctuation in protein level gives rise to
different phenomena in biological system \cite{key-10,key-21,key-22}.
At the macroscopic level, we see that the biological systems are very
much fine-tuned and deterministic. When a cell grows and divide from
its embryonic stage, each and every event occurs at the right time
with certainty. The events are controlled by some proteins since they
are the functional molecules in the cells. Each and every cellular
events are executed by some proteins. They are required for structure,
function, regulation of the body's tissues and organs. Many theoretical
and experimental investigations show that the reduction of protein
level may give rise to different diseases called haploinsufficiency
\cite{key-18,key-130,key-131,key-132,key-133}. In diploid systems,
proteins are produced from two copies of the same gene. If one of
the two copies is mutated, the protein level gets reduced by 50\%.
That reduced amount of proteins are insufficient to carryout their
specific job and gives rise to several diseases called haploinsufficiency.
This shows that for the proper functioning of the proteins, they have
to stay above a critical level. But, there is a chance that the protein
level may fall below the critical level because of the noise present
in it \cite{key-18,key-133,key-134,key-135,key-136,key-137}. Thus,
the noise in protein level has a detrimental role and therefore undesirable.

The biological system or rather living system is basically a non-equilibrium
system. To maintain the state of non-equilibrium in such system, they
need energy from external sources. That is, for the formation of complex
cellular structure and to maintain its activity, living cell has a
cost \cite{key-139,key-140,key-141,key-142,key-143,key-144,key-145,key-146}.
Again, the living system tries to optimize its function by minimizing
the cellular cost through evolution and natural selection \cite{key-142,key-143,key-144}.
The important cellular process like gene expression consists of several
biochemical events e.g., gene activation-inactivation, transcription,
translation, degradation etc., which is not equilibrium rather nonequilibrium
process \cite{key-138,key-140,key-147,key-51,key-52}. Huang et al.
\cite{key-140} studied the fundamental principle of energy consumption
in gene expression. They showed that the speed of stochastic transitions
between the ON and OFF states of the gene is at the cost of energy.
That happens because many regulatory molecules need to accommodate
at the promoter sites to get the ON state of the gene \cite{key-1,key-2,key-3,key-4}
and that requires energy. The fluctuations in the number of regulatory
molecules modulate the stochastic transitions between the ON and OFF
states of a gene. Again, protein synthesis from the ON state of the
gene also requires energy consumption. Study shows that a major part
of the cellular energy is used for amino acid polymerization in protein
synthesis process \cite{key-163}. So, cell consumes energy to maintain
a specific protein level and the consumption amount increases with
the increase of mean protein level and random switching between ON-OFF
states of gene. It is customary to think that there might be some
mechanism of optimization of energy consumption in naturally evolve
complex gene expression process. 

The cellular system achieved a complex structure by evolution and
organised its contents according to its requirement. Now, one can
raise the question: why cellular system has evolved with low copy
number of regulatory molecules and noisy gene expression? Is it a
choice or accident? Does the undesirable noise has any role in the
optimization of energy consumption in gene expression? In this work,
we address such questions related to gene expression and show that
introducing stochastic fluctuations in protein level cell can reduce
the energy consumption efficiently. We show that the fluctuation in
regulatory molecules has a crucial role behind it. We consider a stochastic
model of simple gene regulatory network with two genes, a TF gene
and a functional gene, and study the time evolution of protein's number
from each gene with different noise level in TFs. We observe that
the different amount of noise in TF level determines the noise and
average protein level from the functional gene. We also show from
our stochastic simulation result that high noise in TF level reduces
the cost of energy consumption to keep the protein level above some
critical value from the functional gene rather than the less noisy
TF level. Our general view is that the stochasticity or noise has
a detrimental role in cellular functions as in the electronic system.
But in this study we find that the stochasticity can play the beneficial
role in the cellular functions by saving energy and therefore, may
be a choice of cellular system.

\section{Stochastic model and Analysis}

\subsection{Two-state stochastic model of single gene expression}

In two-state stochastic model of gene expression, a gene can be in
two possible states: ON (active) and OFF (inactive). Genes make random
transitions between the ON and OFF states with specific rate constants.
The protein synthesis takes place in burst from both the states, active
and inactive, of the gene with different rate constants. Let $k_{a}$
and $k_{d}$ are the activation and deactivation rate constants for
the gene. In the active (inactive) state, protein production and degradation
occurs with the rate constants $j_{p}$ ($j_{0}$) and $k_{p}$ respectively.
Here transcription and translation are lumped together into a single
step \cite{key-32,key-161,key-162}. The biochemical steps of gene
expression from a single gene is shown in equation (\ref{eq:1-1})

\begin{equation}
G\:\stackrel{k_{a}}{\longrightarrow}\:G^{*},\;G^{*}\:\stackrel{k_{d}}{\longrightarrow}\:G,\;G\;\stackrel{j_{0}}{\longrightarrow}Pr,\;G^{*}\;\stackrel{j_{P}}{\longrightarrow}Pr,\;Pr\;\stackrel{k_{p}}{\longrightarrow}\varphi\label{eq:1-1}
\end{equation}

Let $p_{1}(n,t)$ ($p_{0}(n,t)$) be the probability that at time
$t,$ gene is in the active (inactive) state $G^{*}$($G$) with $n$
number of protein molecules. The rate of change of probability with
respect to the time is given by the Master equation

\begin{equation}
\frac{\partial p_{0}(n,t)}{\partial t}=k_{d}\,p_{1}(n,t)-k_{a}\,p_{0}(n,t)+j_{0}[p_{0}(n-1,t)-p_{0}(n,t)]+k_{p}[(n+1)p_{0}(n+1,t)-np_{0}(n,t)]\label{eq:A1-1}
\end{equation}

\begin{equation}
\frac{\partial p_{1}(n,t)}{\partial t}=k_{a}\,p_{0}(n,t)-k_{d}\,p_{1}(n,t)+j_{p}[p_{1}(n-1,t)-p_{1}(n,t)]+k_{p}[(n+1)p_{1}(n+1,t)-np_{1}(n,t)]\label{eq:A2-1}
\end{equation}

\begin{flushleft}
For each rate constant, there is a gain term which adds to the probability
and a loss term which subtracts from the probability. The Master Equation
is a rate equation in which probability replaces concentration as
the relevant variable. 
\par\end{flushleft}

We use the standard approach in the theory of stochastic processes
to determine the steady state probability density function \cite{key-165}.
We define the generating functions 
\begin{equation}
F_{0}(z,\,t)=\sum_{n}z^{n}\,p_{0}(n,t),\:\;\;F_{1}(z,\,t)=\sum_{n}z^{n}\,p_{1}(n,t),\;{\textstyle \textrm{and}}\;F(z,t)=\sum_{n}z^{n}\,p(n,t)\label{eq:A3-1}
\end{equation}

where
\begin{equation}
\begin{array}{c}
F(z,t)=F_{0}(z,t)+F_{1}(z,t)\\
p(n,t)=p_{0}(n,t)+p_{1}(n,t)
\end{array}\label{eq:A4-1}
\end{equation}

where $F(z,t)$ and $p(n,t)$ are the total generating function and
total probability density function respectively.

In terms of the generating functions given in equation (\ref{eq:A3-1}),
the Master equations (\ref{eq:A1-1}) and (\ref{eq:A2-1}) can be
written as 
\begin{equation}
\frac{\partial F_{0}(z,t)}{\partial t}=k_{d}F_{1}(z,t)-k_{a}F_{0}(z,t)+j_{0}(z-1)F_{0}(z,t)+k_{p}(1-z)\frac{\partial F_{0}(z,t)}{\partial z}\label{eq:A5-1}
\end{equation}

\begin{equation}
\frac{\partial F_{1}(z,t)}{\partial t}=k_{a}F_{0}(z,t)-k_{d}F_{1}(z,t)+j_{p}(z-1)F_{1}(z,t)+k_{p}(1-z)\frac{\partial F_{1}(z,t)}{\partial z}\label{eq:A6-1}
\end{equation}

\begin{flushleft}
In the steady state (${\displaystyle \frac{\partial F_{0}}{\partial t}}=0$
and ${\displaystyle \frac{\partial F_{1}}{\partial t}}=0)$), adding
equations (\ref{eq:A5-1}) and (\ref{eq:A6-1}) we get
\par\end{flushleft}

\begin{equation}
j_{p}F_{1}(z,t)+j_{0}F_{0}(z,t)=k_{p}\frac{\partial F(z,t)}{\partial z}\label{eq:A7-1}
\end{equation}

\begin{flushleft}
Solving equations (\ref{eq:A4-1}) and (\ref{eq:A7-1}) we get
\par\end{flushleft}

\begin{equation}
F_{1}(z,t)=\frac{k_{p}}{J}\frac{\partial F(z,t)}{\partial z}-\frac{j_{0}F(z,t)}{J}\label{eq:A8-1}
\end{equation}

\begin{equation}
F_{0}(z,t)=\frac{j_{p}F(z,t)}{J}-\frac{k_{p}}{J}\frac{\partial F(z,t)}{\partial z}\label{eq:A9-1}
\end{equation}

where $J=j_{p}-j_{0}.$

\begin{flushleft}
Now, in the steady state, using equations (\ref{eq:A8-1}) and (\ref{eq:A9-1}),
equation (\ref{eq:A5-1}) can be written as 
\par\end{flushleft}

\begin{equation}
(a_{2}z+b_{2})F^{''}(z)+(a_{1}z+b_{1})F^{'}(z)+(a_{0}z+b_{0})F(z)=0\label{eq:A10-1}
\end{equation}

where $a_{2}=1$, $b_{2}=-1$, $a=-(r_{3}+r_{4}),$ $b_{1}=(r_{1}+r_{2}+r_{3}+r_{4})$,
$a_{0}=r_{3}r_{4},$ $b_{0}=-(r_{1}r_{3}+r_{2}r_{4}+r_{3}r_{4}),$
$r_{1}={\displaystyle \frac{k_{a}}{k_{p}}},$ $r_{2}={\displaystyle \frac{k_{d}}{k_{p}}}$,
$r_{3}={\displaystyle \frac{j_{p}}{k_{p}}}$ and $r_{4}=\frac{{\displaystyle j_{0}}}{{\displaystyle k_{p}}}$.

\begin{flushleft}
The exact solution of equation (\ref{eq:A10-1}) is given by (using
Mathematica)
\par\end{flushleft}

\begin{equation}
F(z)=N\:e^{K\,z}\,_{1}F_{1}(a_{3};b_{3};(\frac{z-\mu}{\lambda}))\label{eq:A11-1}
\end{equation}

where $\mu=-{\textstyle \nicefrac{{\displaystyle b_{2}}}{{\displaystyle a_{2}}}}$,
$K=\frac{\sqrt{D}-a_{1}}{2\,a_{2}}$, $D={\displaystyle a_{1}^{2}-4a_{0}a_{2}}$,
$a_{3}=\frac{{\displaystyle b_{2}K^{2}+b_{1}K+b_{0}}}{2\,a_{2}K+a_{1}}$,
$b_{3}=(a_{2}b_{1}-a_{1}b_{2})a_{2}^{-1}$ and $\lambda=-{\displaystyle \frac{a_{2}}{2\,a_{2}K+a_{1}}}$.
$_{1}F_{1}(a;b;z)$ is the confluent hypergeometric function and $N$
is the normalization constant. $N$ is determined from the condition
$F(1)=1$ and is given by $N=\{e^{K}\:F_{1}(a_{3};b_{3};(\frac{1-\mu}{\lambda}))\}^{-1}$.

\begin{flushleft}
Differentiating equations (\ref{eq:A3-1}) and (\ref{eq:A11-1}) $n$
times w.r.t. $z$ at $z=0$ and then comparing both sides we have
the total probability density function
\par\end{flushleft}

\begin{equation}
p(n)=N\sum_{m=0}^{n}\frac{K^{n-m}\,(1/\lambda)^{m}\,\Gamma(a_{3}+m)\,\Gamma(b_{3})}{(n-m)!\,m!\,\Gamma(a_{3})\,\Gamma(b_{3}+m)}\,_{1}F_{1}(a_{3}+m;\,b_{3}+m;\,-(1/\lambda))\label{eq:A12-1}
\end{equation}

Fig. 1 shows the distribution of proteins ($p(n)$) versus number
of proteins ($n$) plot for $r_{3}=500$, $r_{4}=50$ and for $r_{1},r_{2}>1$.
Fig. 2 shows the same plot but with different values of $r_{1}$ and
$r_{2}$ with $r_{1},r_{2}<1$. It is seen that the distribution is
bimodal (Fig. 2) for $r_{1},r_{2}<1$ and unimodal (Fig. 1) for $r_{1},\:r_{2}>1$.
For $r_{1},r_{2}=1$, the distribution becomes uniform. The unimodal
responses can also be obtained either for $r_{1}>1$ or $r_{2}>1$
only with mode towards higher value or lower value respectively. The
rate constants $r_{3}$ and $r_{4}$ determine the positions of the
upper and lower modes of the bimodal distribution respectively.

\noindent Using equation (\ref{eq:A10-1}) one can easily derive the
expression for mean ($<n>$) and variance ($var$) and are given by

\begin{equation}
<n>=\frac{r_{1}}{r_{1}+r_{2}}r_{3}+\frac{r_{2}}{r_{1}+r_{2}}r_{4}\label{eq:A14-1}
\end{equation}

\begin{equation}
var=<n>(1+\frac{r_{1}r_{2}(r_{3}-r_{4})^{2}}{(r_{1}+r_{2})(r_{1}+r_{2}+1)(r_{1}r_{3}+r_{2}r_{4})})\label{eq:A15-1}
\end{equation}

In equation (\ref{eq:A15-1}), the first term appears due to the random
birth and death of proteins and the second term appears due to the
random transitions between the ON and OFF states of the gene. When
$r_{1}$ and $r_{2}$ are very high i.e., the number of transitions
between ON and OFF states of the gene is very large \cite{key-149},
the noise in protein level about the mean is very low. Now, as $r_{1}$
and $r_{2}$ are decreases, the number of transitions between ON and
OFF states of the gene also decreases, the noise or fluctuation about
mean level increases. The width of the distributions correctly reflects
that in Fig. 1(a) and 1(b). The graded responses of proteins are always
less noisy compared to the binary responses for fixed average level
of proteins. Sometimes, mean independent fluctuation or noise is measured
by a quantity called Fano Factor (FF). The Fano Factor is defined
by $var/<n>$ \cite{key-5,key-7}. Fig. 3 shows the variation of Fano
Factor with $r_{1}$ and $r_{2}$ for fixed value of $r_{3}$ and
$r_{4}$. Fano Factor increases as $r_{1}$ and $r_{2}$ are decreases.

\begin{figure}
\begin{centering}
\includegraphics[width=5cm,height=3cm]{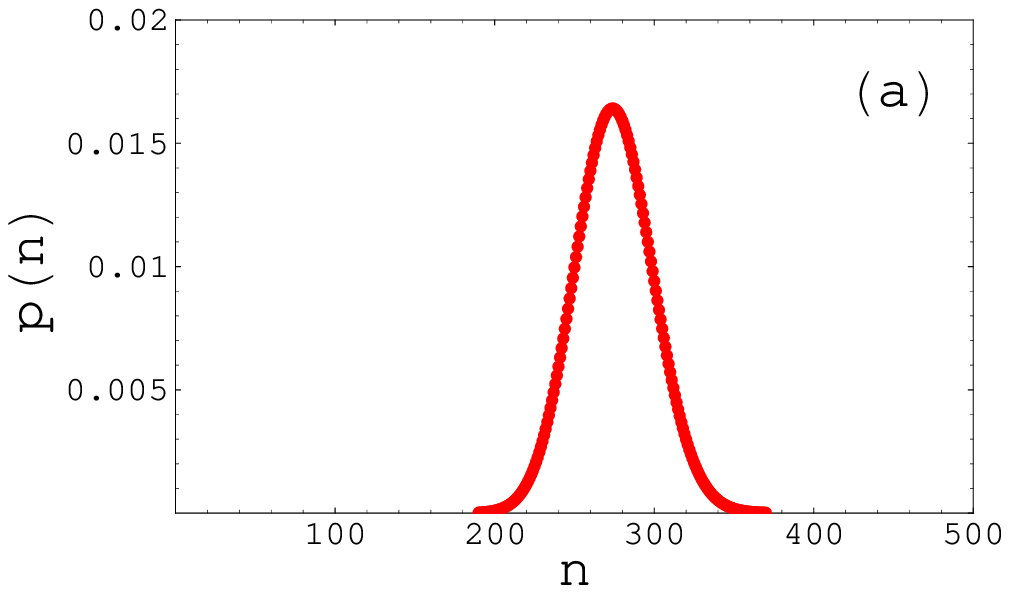} \includegraphics[width=5cm,height=3cm]{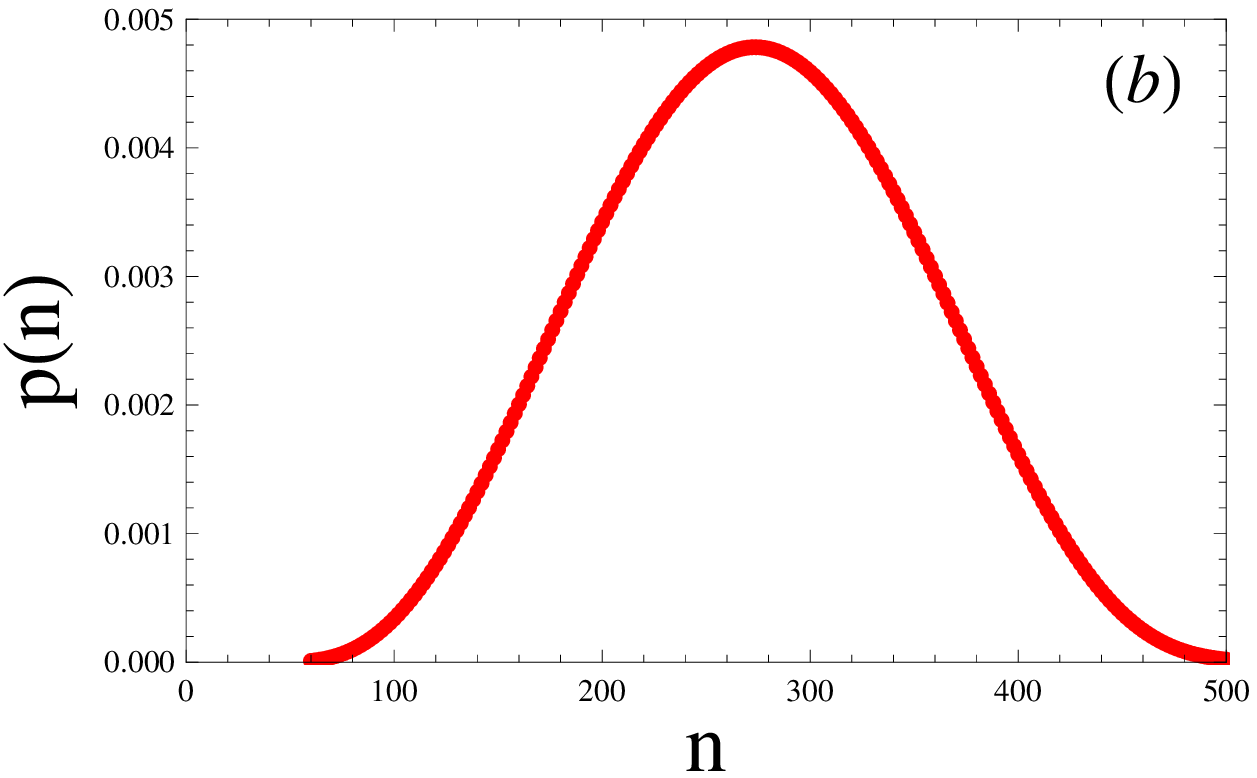}
\par\end{centering}
Fig. 1. Unimodal responses. In Fig. (a) the rate constants are $k_{a}=80.0$,
$k_{d}=80.0$, $J_{p}=500.0$, $J_{0}=50.0$, $k_{p}=1.0$. In Fig.
(b) $k_{a}=4.0$, $k_{d}=4.0$ and remaining are the same. As the
value of rate constants $k_{a}$ and $k_{d}$ decreases the width
of the distribution increases. Equal values of $k_{a}$ and $k_{d}$
makes the distribution symmetric about the mode.
\end{figure}

\begin{figure}
\begin{centering}
\includegraphics[width=5cm,height=3cm]{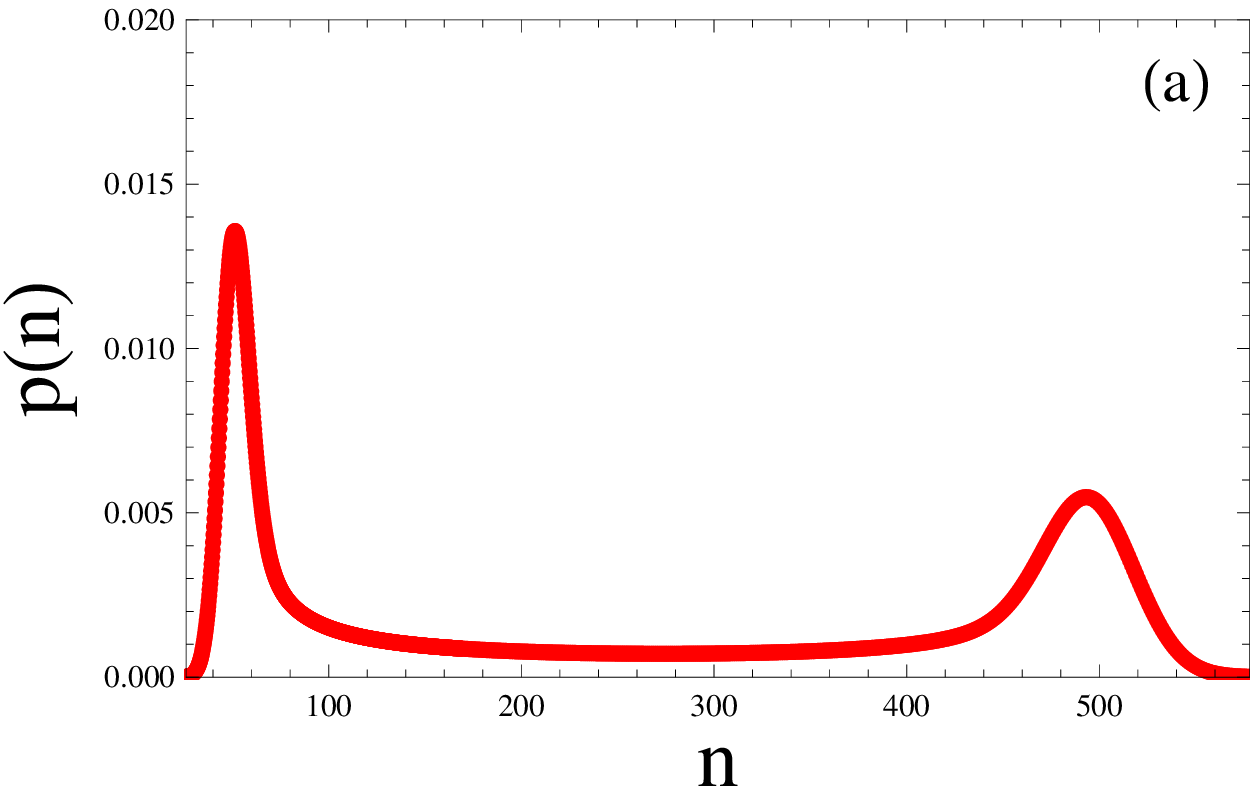}
\includegraphics[width=5cm,height=3cm]{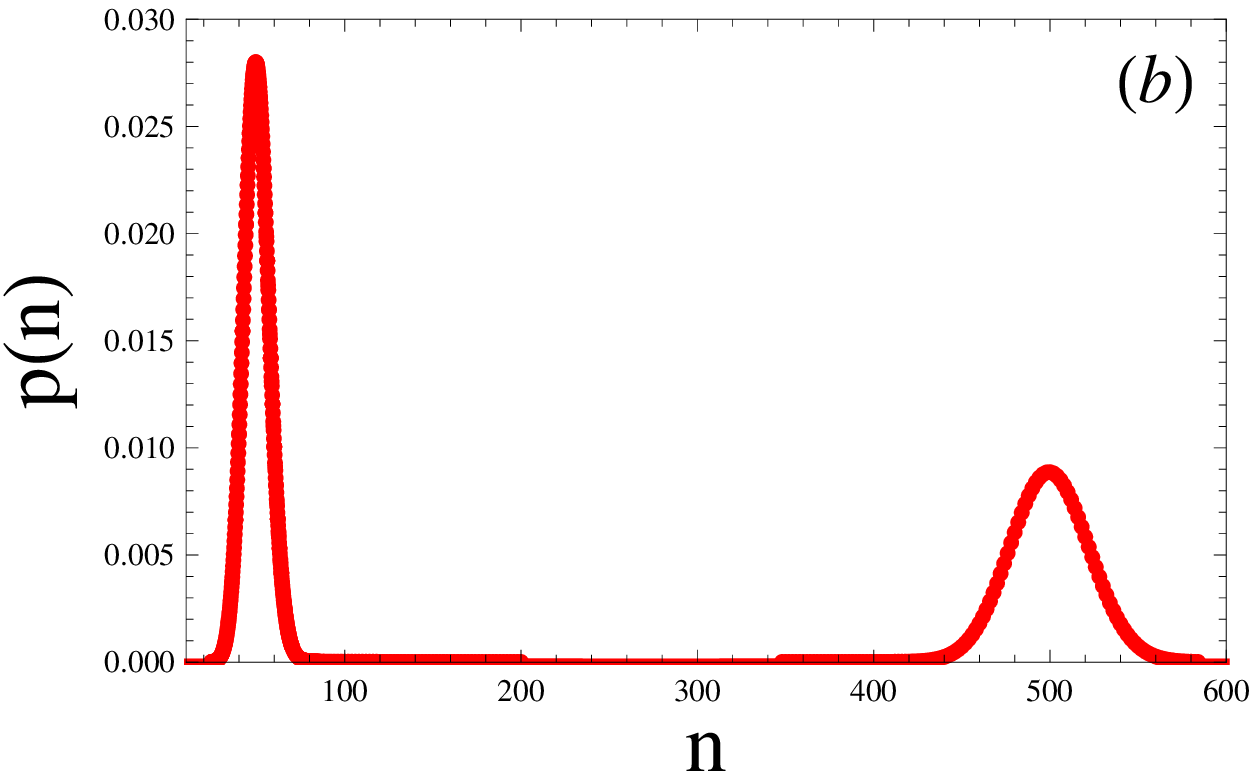}
\par\end{centering}
Fig. 2. Bimodal responses. In Fig. (a) the rate constants are $k_{a}=0.2$,
$k_{d}=0.2$, $J_{p}=500.0$, $J_{0}=50.0$, $k_{p}=1.0$. In Fig.
(b) $k_{a}=0.002$, $k_{d}=0.002$ and remaining are the same. For
the bimodal responses, the protein level fluctuates between low and
high level among with the noise due to random birth and death of the
proteins in the respective levels.
\end{figure}

\begin{figure}
\begin{centering}
\includegraphics[width=7cm,height=5cm]{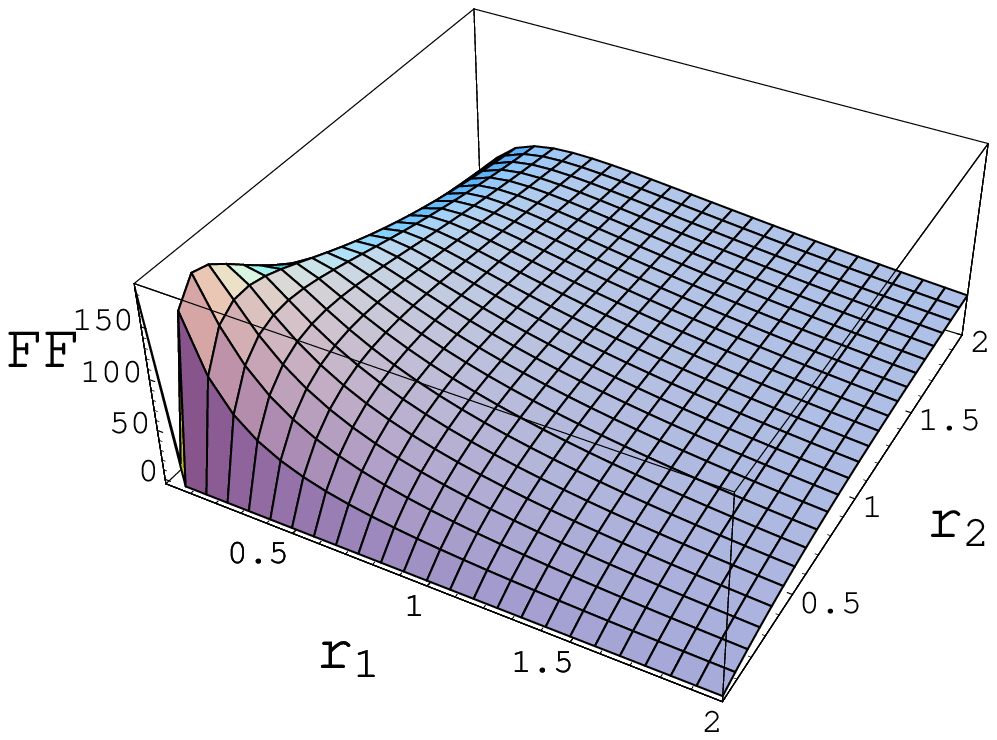}
\par\end{centering}
Fig. 3. Variation of Fano Factor (FF) with $r_{1}$ and $r_{2}$ for
$r_{3}=500$ and $r_{4}=50.$
\end{figure}

The conditions of unimodal (either $r_{1}>1$ or $r_{2}>1$ or $r_{1},\,r_{2}>1$)
and bimodal ($r_{1},\,r_{2}<1$) responses are actually given in Ref.\cite{key-13}
with approximate solution of probability density function for protein
number. The bimodal distribution of protein level is also known as
all-or-none phenomena in cellular system and can be observed without
any feedback processes also \cite{key-13,key-30}. It can be shown
that the gene expression response depends on the relative values of
the parameters rather than the absolute values. If all the rate constants
are multiplied by the same factor, the distribution will remain unchangeed.

\subsection{Stochastic model of two-gene network}

In the two genes model, we consider a simple gene regulatory network
consisting of two genes: transcription factor (TF) gene and functional
gene (Fig. 4). The proteins from the TF gene activate the protein
synthesis from the functional gene \cite{key-137}. The proteins from
the functional gene execute some important functions in the cell as
$G6PC$ gene in liver \cite{key-12}. Each gene of the network follows
the basic biochemical steps considered in Section 2.1. The steps are
shown in equations (\ref{eq:1}) and (\ref{eq:2}) along with the
rate constant for the respective reaction. We also assume that the
activation of functional gene requires $n$ number of TFs. That $n$
TF molecules bind the promoter sites of the functional gene through
$n$ steps to activate the functional gene. The $n$ steps ($n$=1,
2, 3, 4 etc.) activation process of functional gene by TFs can be
mapped by the Hill function and can be represented as single step
process \cite{key-132,key-133,key-137,key-53}. This is shown in equation
(\ref{eq:2}). 

\begin{figure}
\begin{centering}
\includegraphics[width=5cm,height=1.5cm]{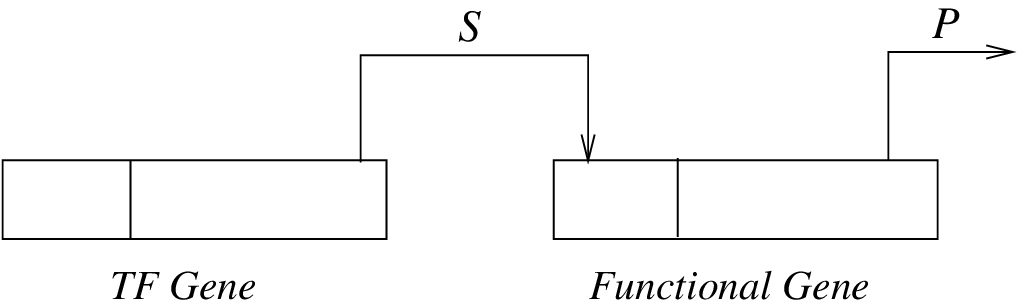}
\par\end{centering}
Fig. 4. Schematic diagram of gene network consisting of two genes:
TF gene and Functional gene, Protein $S$ (TF) from TF gene is regulating
the synthesis of functional protein $P$ from functional gene.
\end{figure}

The biochemical reactions are considered as follows:

\vspace{0.1cm}

For TF gene:

\begin{equation}
G_{T}\:\stackrel{k_{aT}}{\longrightarrow}\:G_{T}^{*},\;G_{T}^{*}\:\stackrel{k_{dT}}{\longrightarrow}\:G_{T},\;G_{T}\;\stackrel{J_{0T}}{\longrightarrow}S,\;G_{T}^{*}\;\stackrel{J_{PT}}{\longrightarrow}S,\;S\;\stackrel{k_{pT}}{\longrightarrow}\varphi\label{eq:1}
\end{equation}

\vspace{0.1cm}

For Functional gene: 

\begin{equation}
G_{F}\:\stackrel{k_{aF}^{|}}{\longrightarrow}\:G_{F}^{*},\;G_{F}^{*}\:\stackrel{k_{dF}}{\longrightarrow}\:G_{F},\;G_{F}\;\stackrel{J_{0F}}{\longrightarrow}P,\;G_{F}^{*}\;\stackrel{J_{PF}}{\longrightarrow}P,\;P\;\stackrel{k_{pF}}{\longrightarrow}\varphi\label{eq:2}
\end{equation}

\vspace{0.2cm}

The genes can be in two possible states : OFF ($G_{i}$) or ON ($G_{i}^{*}$)
($i=T$or F). Protein synthesis takes place from the ON state of the
gene with higher rate ($J_{Pi}$) than that from the OFF state ($J_{0i}$).
Both the proteins have some degradation rate constant $k_{pi}$. $k_{aF}^{/}$
($k_{aT}$) and $k_{dF}$ ($k_{dT}$) are the activation and deactivation
rate constants respectively for the functional gene (TF gene). The
activation rate constant for the functional gene is given by $k_{aF}^{/}=k_{aF\:}f$,
where $f$ is the Hill function and is given by $f=\frac{(S/K)^{n}}{1+(S/K)^{n}}$
\cite{key-137,key-52}. $S$ is the TF number and at $K=S$, the Hill
function is $f=0.5$, $n$ is the Hill coefficient. The Hill function
is a nonlinear sigmoidal shape for $n\geq2$. A small fluctuations
in TF numbers about $S=K$ gives rise to large fluctuations in $k_{aF}^{/}$
\cite{key-132}. The expression for mean TF level ($<S>$) and mean
functional protein level ($<P>$) is given by equations (\ref{eq:3})
and (\ref{eq:4}) respectively. The mean functional protein level
depends on the instantaneous value of TF number since $k_{aF}^{/}$
depends on $S$.

\begin{equation}
S_{mean}=<S>=(\frac{k_{aT}}{k_{aT}+k_{dT}})\frac{J_{pT}}{k_{pT}}+(\frac{k_{dT}}{k_{aT}+k_{dT}})\frac{J_{0T}}{k_{pT}}\label{eq:3}
\end{equation}

\begin{equation}
P_{mean}=<P>=(\frac{k_{aF}^{/}}{k_{aF}^{/}+k_{dF}})\frac{J_{pF}}{k_{pF}}+(\frac{k_{dF}}{k_{aF}^{/}+k_{dF}})\frac{J_{0F}}{k_{pF}}\label{eq:4}
\end{equation}

\subsection{Stochastic simulation, results and analysis}

We simulate the biochemical processes of gene expression using Gillespie
algorithm \cite{key-148}. The rate constants of different biochemical
steps in gene expression determine the dynamics of gene expression.
We choose the protein synthesis and degradation rate constants for
TF gene as $J_{pT}=500.0$, $J_{0T}=50.0$, $k_{pT}=1.0$. The choice
of the value of rate constants $J_{pT}$ and $J_{0T}$ is arbitrary
and can be chosen any value for the study. For chosen value and for
$k_{aT}$ = $k_{dT}$, the mean TF level is $275$. We varied the
noise in TFs level keeping $S_{mean}$ fixed by varying $k_{aT}$
and $k_{dT}$ from a very high value (low noise) to a very low value
(large noise) to observe the impact of noise of TF on the functional
protein level. We divide wide region of parameters space for $k_{aT}$
and $k_{dT}$ into four different regions with different noise profiles
of TF gene (Fig. 3) and responses from functional gene. We call them
four different major Strategies. They are chosen as: \noun{Strategy
I}: Low noise in TF level ($\frac{k_{aT}}{k_{pT}}$ and $\frac{k_{dT}}{k_{pT}}$
>\textcompwordmark > 10), \noun{Strategy} II: Moderate noise in TF
level (1<$\frac{k_{aT}}{k_{pT}}$ and $\frac{k_{dT}}{k_{pT}}$<10
), \noun{Strategy} III: High noise in TF level (0.1<$\frac{k_{aT}}{k_{pT}}$
and $\frac{k_{dT}}{k_{pT}}$ < 1), \noun{Strategy} IV: Very high noise
in TF level ($\frac{k_{aT}}{k_{pT}}$ and $\frac{k_{dT}}{k_{pT}}$
<\textcompwordmark <0.1). The functional proteins do some important
job and therefore should not degrade too early after synthesis and
the noise should be small in it. So the rate constants are chosen
as: $k_{aF}^{/}=k_{aF\:}f$, $K=S_{mean}$, $k_{aF}=8.0$, $k_{dF}=4.0$,
$J_{pF}=5.0$, $J_{0F}=0.5$, $k_{pF}=0.005$. That gives noisy graded
protein level for large TF numbers ($S>K$). For the assumption $k_{aF}=2k_{dF}$,
the fluctuating $S$ with $K=S_{mean}$ gives $k_{aF}^{/}=k_{dF}$
\cite{key-137}. The important point is that the protein synthesis
and degradation dynamics for functional gene is assumed to be slower
than the TF gene. The rate constants chosen here are almost similar
to the study of Kaern et al. \cite{key-10}. The above values of the
rate constants give $P_{mean}$= $550$ for $S=S_{mean}=275$ and
$K=S_{mean}$ and are kept fixed throughout the study. As already
discussed, the $50\%$ reduction in protein number for a gene can
create problems (haploinsufficiency) in its functioning. This suggest
that protein level must lie above a critical or threshold level for
proper execution of its task. But there is no study showing the accurate
value of the critical level. In our study, we choose $60\%$ of the
value of $P_{mean}=550$ is the critical value i.e.,$P_{crt}=330$,
for the functioning of the protein from functional gene. The estimation
of energy consumption for the gene expression of two-gene network
requires the exact amount of energy cost for the each steps in equations
(\ref{eq:1}) and (\ref{eq:2}). But, the exact value of energy cost
per transition from OFF to ON state of a gene is not known. We can
only say that to make transitions from OFF to ON state, energy consumption
is essential for the assembling of different regulatory molecules
at the promoter sites and it increases when number of transitions
increases \cite{key-140}. That transition is determined by the activation
and deactivation rate constants $k_{aT}$ and $k_{dT}$ ($k_{aF}^{/}$,
$k_{dF}$) for the TF gene (functional gene). Similarly, the exact
value of energy cost for the synthesis of proteins per unit average
value is not known. We can only say that as the mean protein level
increases the energy consumption also increases \cite{key-140,key-163}.
In our simulation study, we have noted the number of OFF to ON state
transitions for TF ($n_{T}$) and functional gene ($n_{F}$) and then
presented an approximate calculation of energy cost for the network
over a fixed time at the steady state. That helps us to compare the
energy consumption of gene expression for different strategies. The
simulation runs for $t=2000$ units for the evolution of proteins
and the steady state is considered at $t\geqslant800.$ 
\begin{figure}
\begin{centering}
\includegraphics[width=8cm,height=5.5cm]{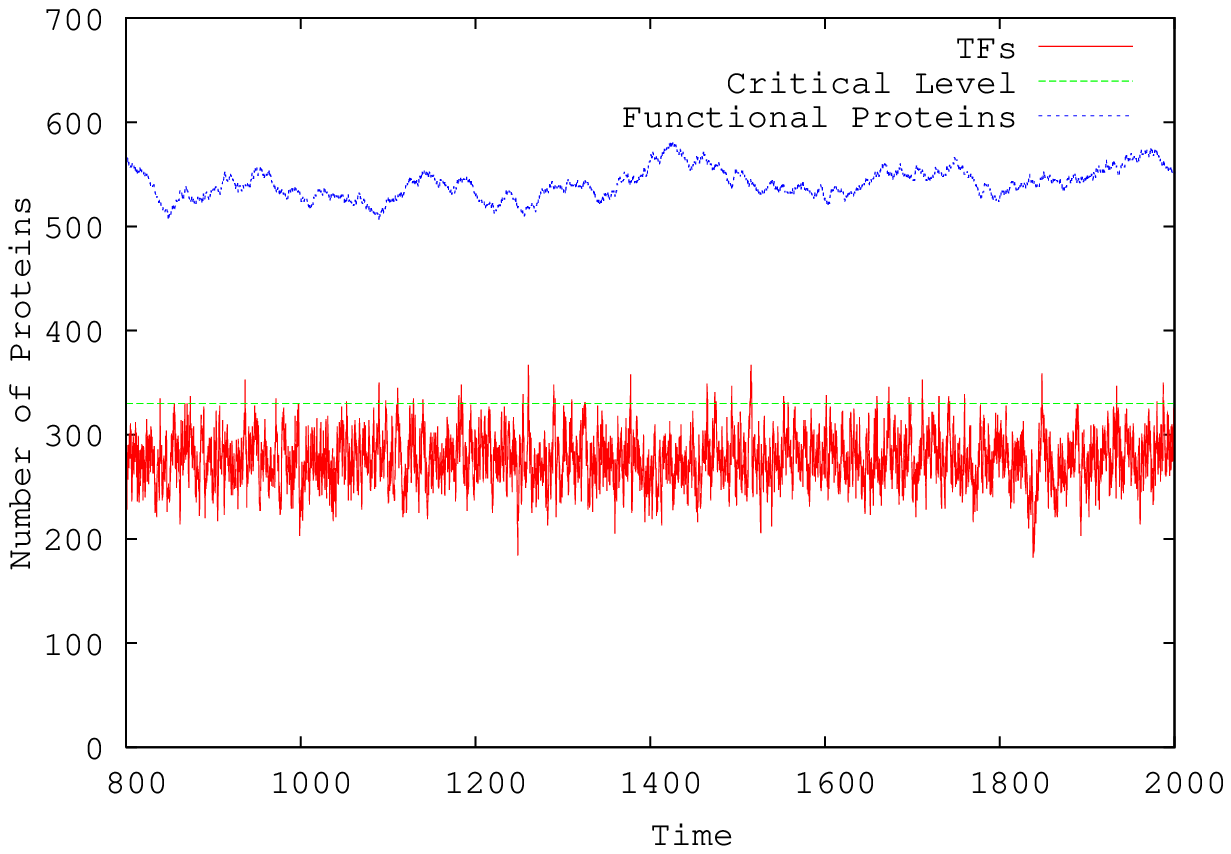}\includegraphics[width=5cm,height=6cm]{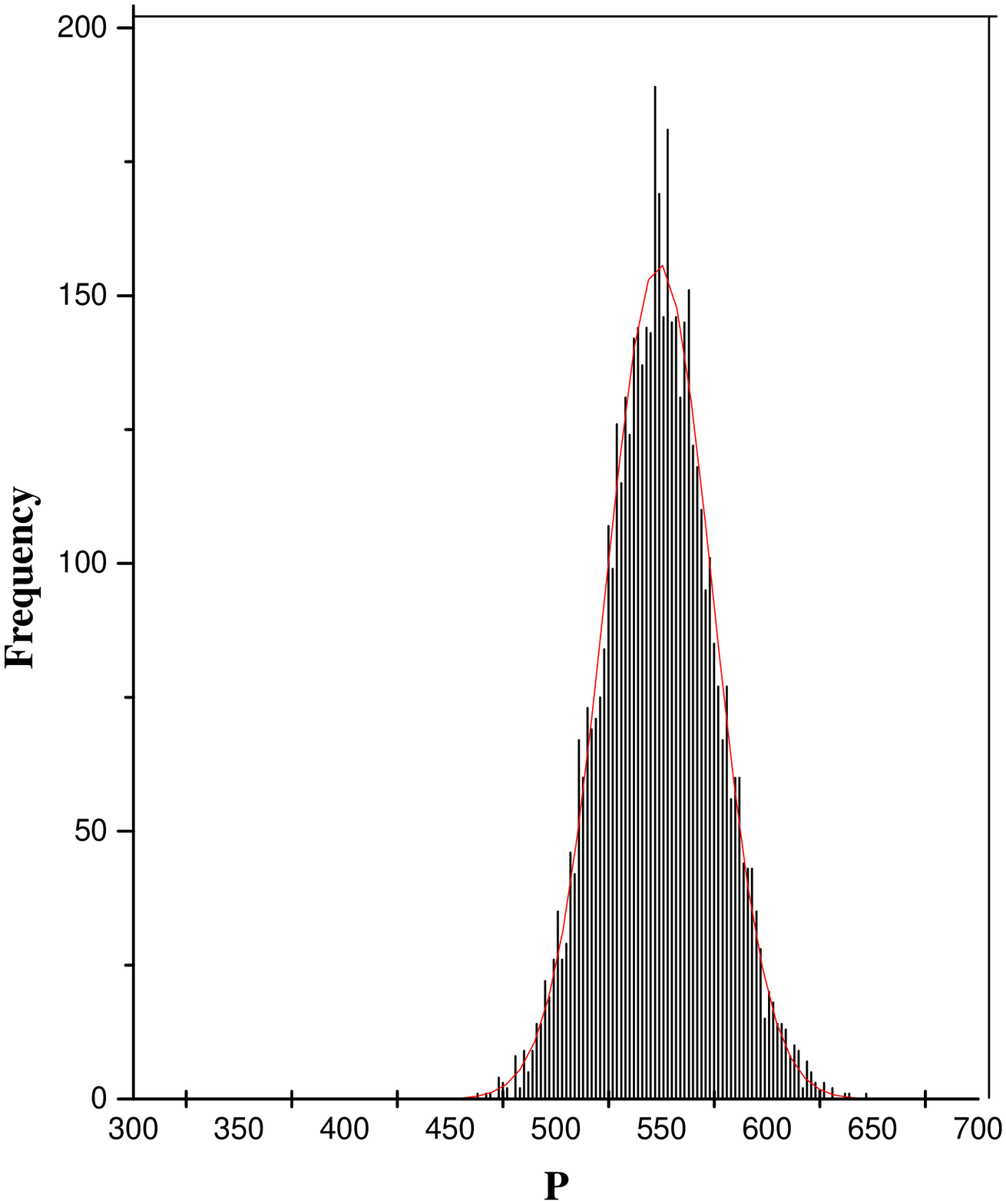}
\par\end{centering}
$\qquad\qquad\qquad\qquad\qquad\qquad\qquad(a)\qquad\qquad\qquad\qquad\qquad\qquad\qquad\qquad\qquad(b)\qquad\qquad$

Fig. 5. (a) Evolution of protein molecules corresponding to \noun{Strategy
I} (at the steady state). For the TFs (red solid line), the rate constants
are $k_{aT}=80.0$, $k_{dT}=80.0$, $J_{pT}=500.0$, $J_{0T}=50.0$,
$k_{pT}=1.0$. For the functional proteins (blue dotted line), the
rate constants are $k_{aF}=8.0$, $k_{dF}=4.0$, $J_{pF}=5.0$, $J_{0F}=0.5$,
$k_{pF}=0.005$, $n=4$ and $K=S_{mean}=275$. The functional protein
level is well above the critical value (green dashed line). (b) Histogram
for functional proteins at the steady state. The Goussian fit gives
$<P>=548$ and Standard Deviation = $25.44.$ 
\end{figure}

\noun{Strategy I:} The random transitions between the active and inactive
states of the TF gene is very fast with respect to TF degradation
rate ($k_{aT}=80.0$, $k_{dT}=80.0$, $k_{pT}=1.0$ ). The time evolution
of TFs and functional proteins are shown in Fig. 5(a). Protein level
from TF gene is now unimodal in nature with less noisy expression
level with mean value at $275$ (Fig. 1(a)). Because of the less noisy
level of regulatory molecules, low fluctuating functional protein
level (with the steady state average value $<p>=548$ and Standard
Deviation (SD) = $25.44$ (Fig. 5(b))) arises from the functional
gene and that lies always above the critical value ($330$) shown
by a dash-dot line in figure (Fig. 5(a)). The number of transitions
between inactive to active state $n_{T}$ is $47842$ ($n_{F}=2356$).
Let us consider the energy cost for each transition from inactive
to active state of both the genes is approximately \emph{H} units.
Let us also assume that the approximate average energy cost to produce
per unit mean protein level from functional gene is \emph{K} units.
Therefore, the total energy cost in \noun{Strategy I} is $E_{1}$$=A+(n_{T}$+$n_{F})H$
+$<p>K$ $=A+50198H+548K$. Here, $A$ is the average energy cost
for all other processes in protein synthesis of the two-gene network. 

\begin{figure}
\begin{centering}
\includegraphics[width=8cm,height=5.5cm]{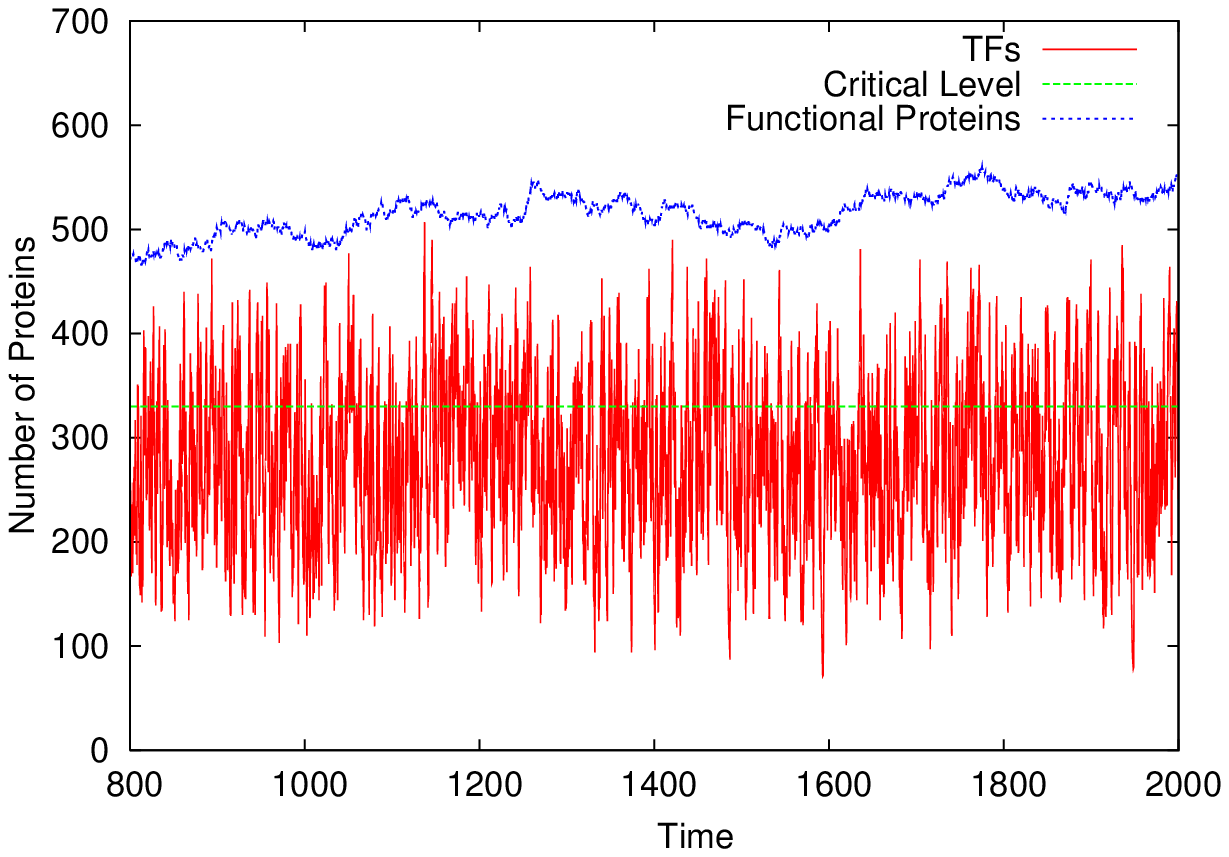}\includegraphics[width=5cm,height=6cm]{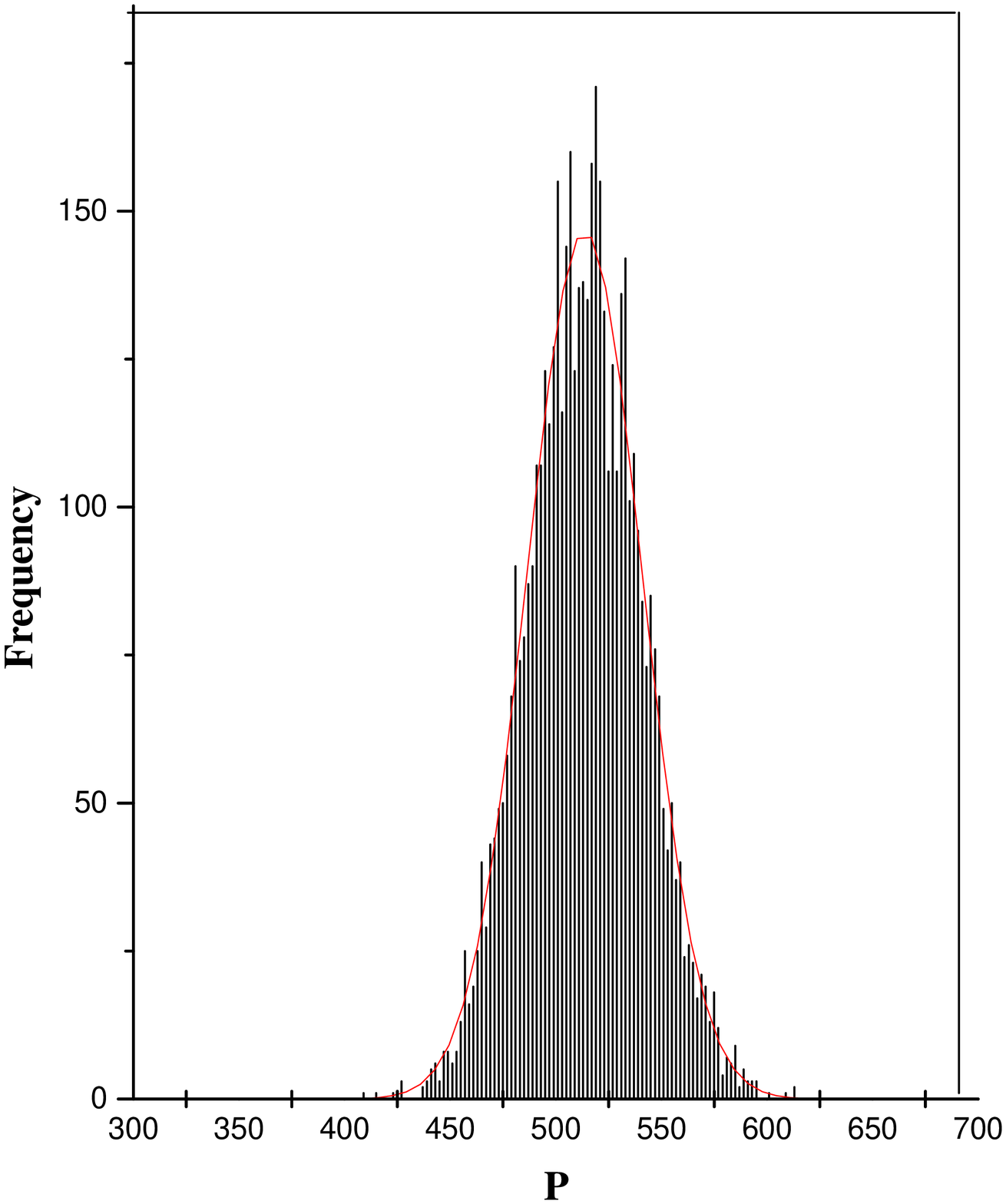}
\par\end{centering}
$\qquad\qquad\qquad\qquad\qquad\qquad\qquad(a)\qquad\qquad\qquad\qquad\qquad\qquad\qquad(b)\qquad\qquad$

Fig. 6. (a) Evolution of protein molecules corresponding to \noun{Strategy
I}I (at the steady state). For the TFs (red solid line), the rate
constants are $k_{aT}=4.0$, $k_{dT}=4.0$, $J_{pT}=500.0$, $J_{0T}=50.0$,
$k_{pT}=1.0$. For the functional proteins (blue dotted line), the
rate constants are $k_{aF}=8.0$, $k_{dF}=4.0$, $J_{pF}=5.0$, $J_{0F}=0.5$,
$k_{pF}=0.005$, $n=4$ and $K=S_{mean}=275$. The functional protein
level is well above the critical value (green dashed line). (b) Histogram
for functional proteins at the steady state. The Goussian fit gives
$<P>=513$ and Standard Deviation = $27.25.$ 
\end{figure}
\begin{figure}
\begin{centering}
\includegraphics[width=8cm,height=5.5cm]{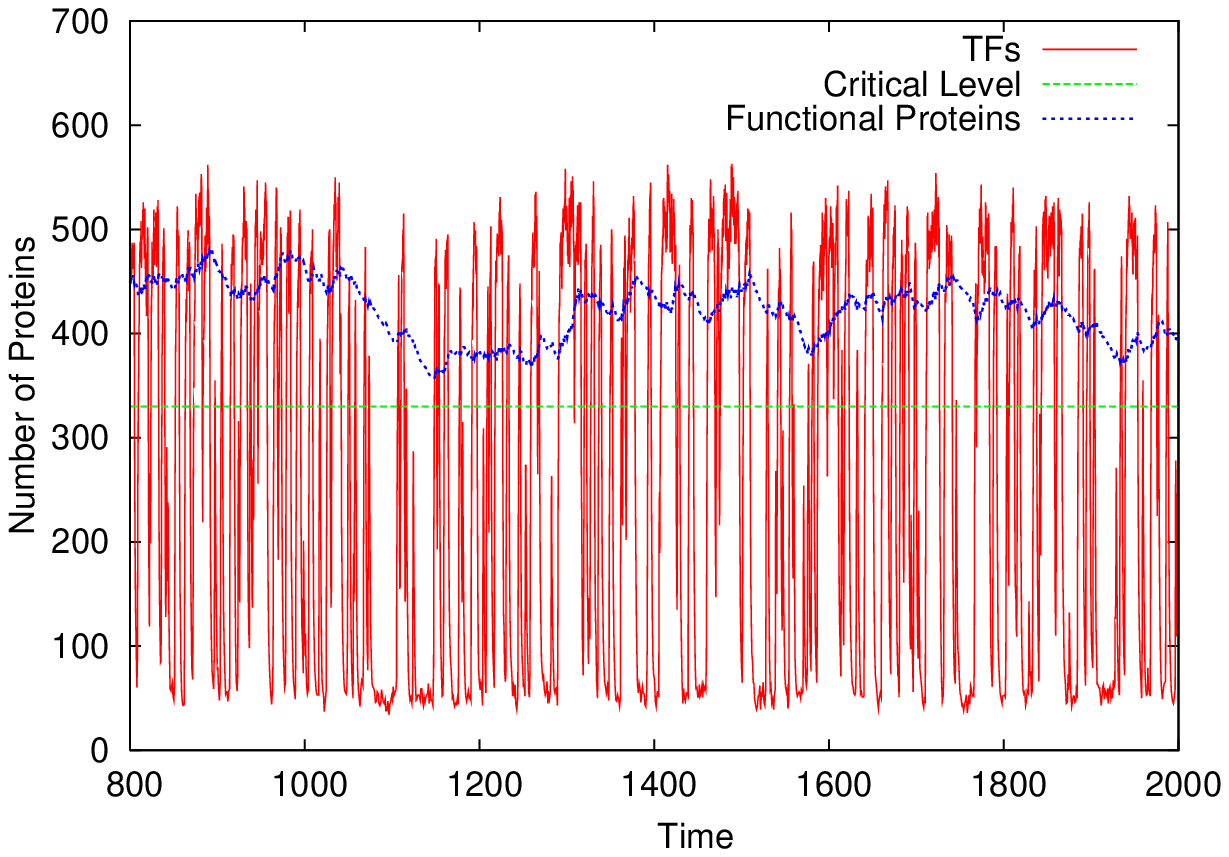}\includegraphics[width=5cm,height=6cm]{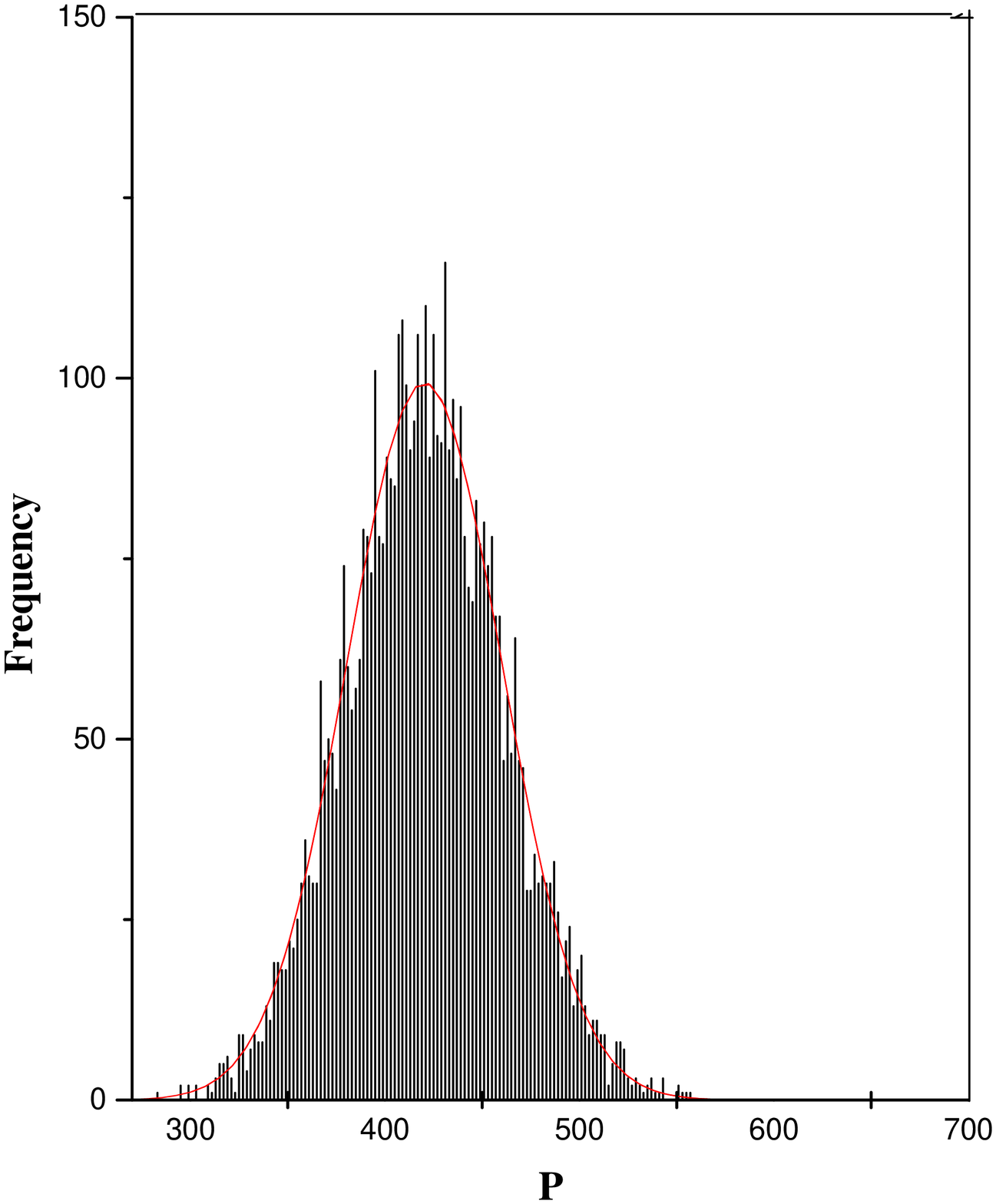}
\par\end{centering}
$\qquad\qquad\qquad\qquad\qquad\qquad\qquad(a)\qquad\qquad\qquad\qquad\qquad\qquad\qquad\qquad\qquad(b)\qquad\qquad$

Fig. 7. (a) Evolution of protein molecules corresponding to \noun{Strategy
III} (at the steady state). For the TFs (red solid line), the rate
constants are $k_{aT}=0.2$, $k_{dT}=0.2$, $J_{pT}=500.0$, $J_{0T}=50.0$,
$k_{pT}=1.0$. For the functional proteins (blue dotted line), the
rate constants are $k_{aF}=8.0$, $k_{dF}=4.0$, $J_{pF}=5.0$, $J_{0F}=0.5$,
$k_{pF}=0.005$, $n=4$ and $K=S_{mean}=275$. The functional protein
level is well above the critical value (green dashed line). (b) Histogram
for the functional proteins at the steady state. The Goussian fit
gives $<P>=420$ and Standard Deviation = $40.05.$ 
\end{figure}

\noun{Strategy II:} Here, the random transitions between the active
and inactive states of TF gene is moderate with respect to the degradation
rate ( $k_{aT}=4.0$, $k_{dT}=4.0$, $k_{pT}=1.0$) of TF proteins.
The time evolution of TFs and functional proteins are shown in Fig.
6(a). The unimodal response of TFs is shown in Fig. 1(b). We got the
value of $n_{T}$ =$2376$ ($n_{F}=2206$) with low fluctuating functional
protein level (with the steady state average value $<p>=513$ and
SD = $27.25$ (Fig. 6(b))). The total energy consumption is $E_{2}=A+4582H+513K$.
It is seen $E_{2}<E_{1}$ because of the lower number of transitions
to active states of both the genes and a bit low value of mean protein
level from functional gene. In \noun{Strategy II} (Fig.6), the noise
in TF protein level and also in functional protein level is greater
than that in \noun{Strategy I} (Fig. 5). 
\begin{figure}
\begin{centering}
\includegraphics[width=8cm,height=5.5cm]{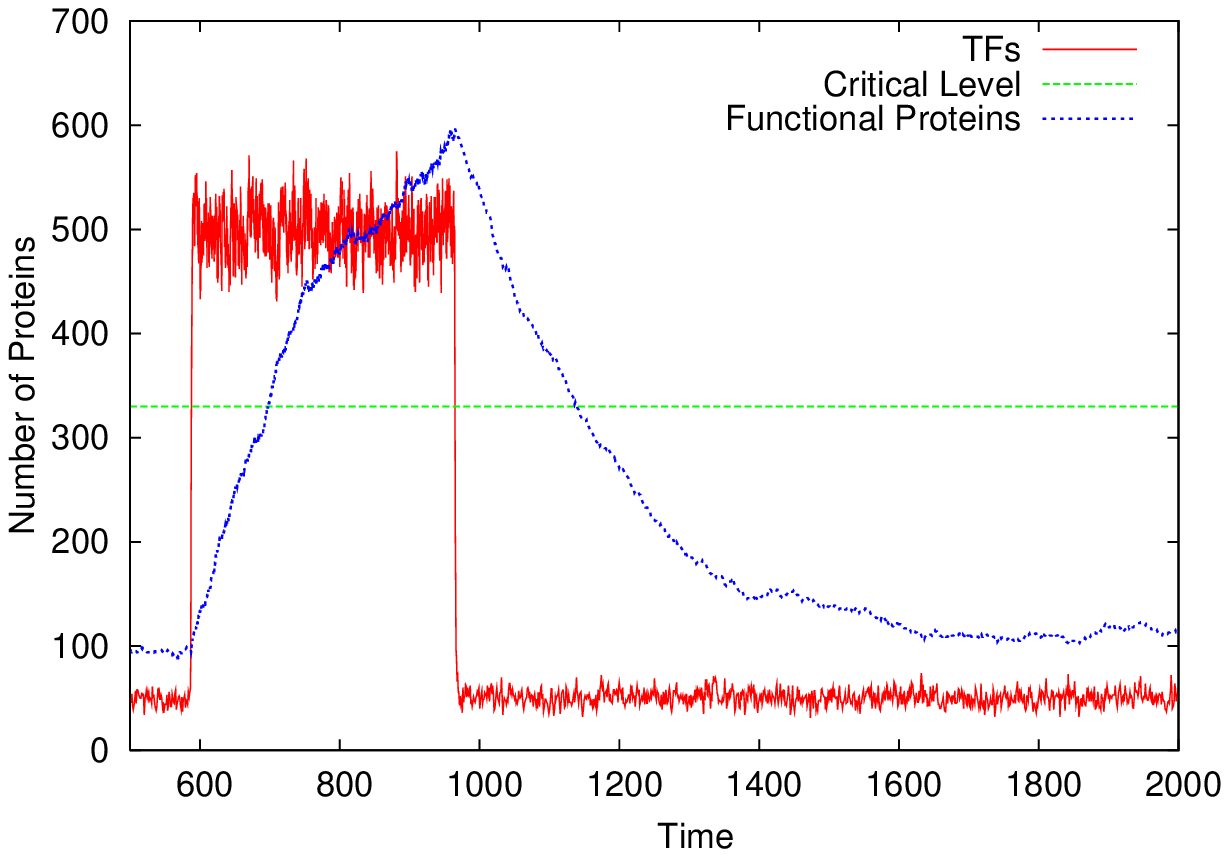}\includegraphics[width=5cm,height=6cm]{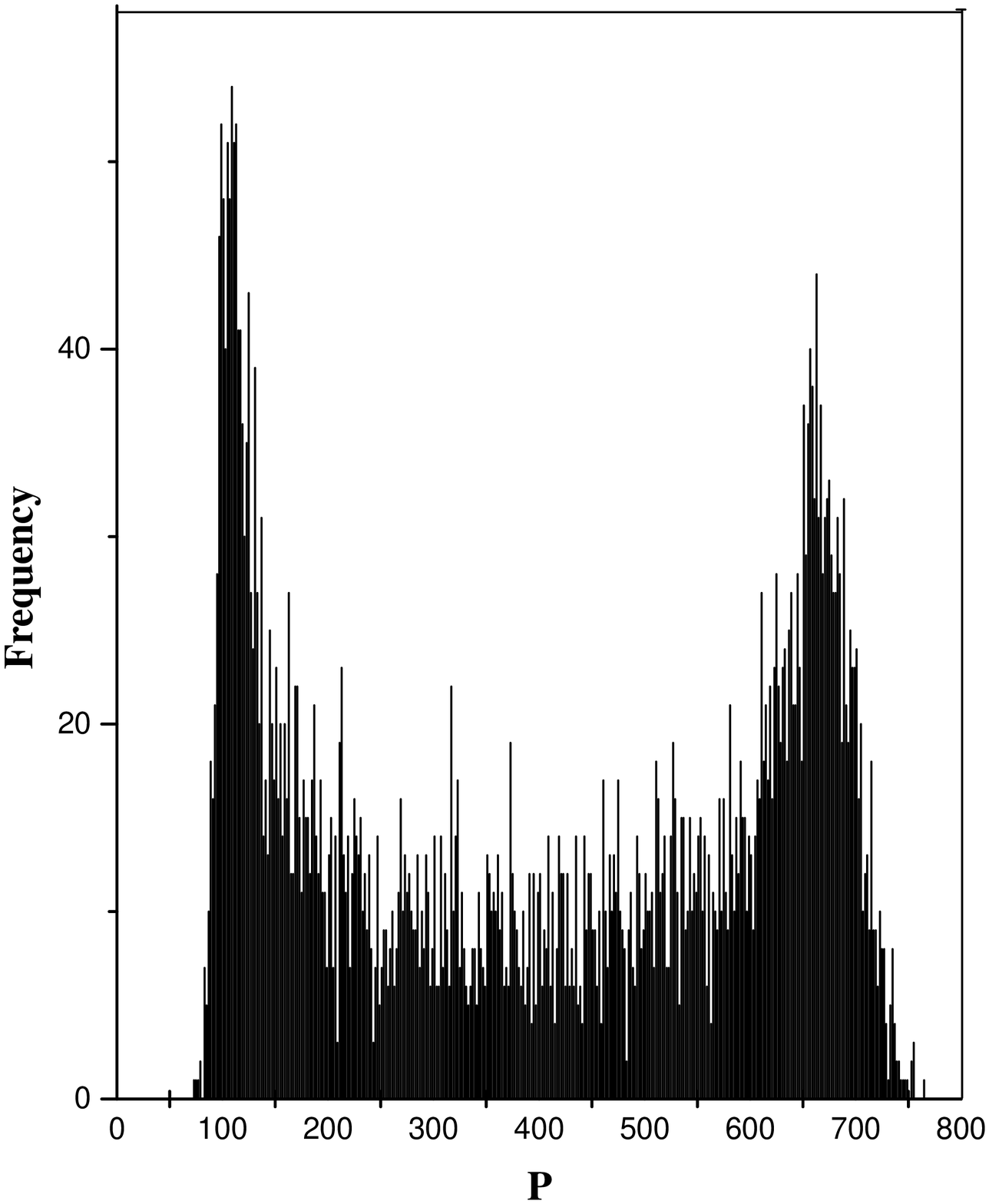}
\par\end{centering}
$\qquad\qquad\qquad\qquad\qquad\qquad\qquad(a)\qquad\qquad\qquad\qquad\qquad\qquad\qquad\qquad\qquad(b)\qquad\qquad$

Fig. 8. (a) Evolution of protein molecules corresponding to \noun{Strategy
I}V (at the steady state). For the TFs (red solid line), the rate
constants are $k_{aT}=0.002$, $k_{dT}=0.002$, $J_{pT}=500.0$, $J_{0T}=50.0$,
$k_{pT}=1.0$. For the functional proteins (blue dotted line), the
rate constants are $k_{aF}=8.0$, $k_{dF}=4.0$, $J_{pF}=5.0$, $J_{0F}=0.5$,
$k_{pF}=0.005$, $n=4$, and $K=S_{mean}=275$. The functional protein
level falls below the critical level (green dashed line) (b) Histogram
for functional proteins at the steady state. The Histogram is fitted
with a bimodal distribution which gives $<P>=350$ and $SD=187.2.$
\end{figure}

\noun{Strategy III:} We consider here the slow transition rate constants
between the active and inactive states of TF gene than the degradation
rate constant ($k_{aT}=0.2$, $k_{dT}=0.2$, $k_{pT}=1.0$). The time
evolution of TFs and functional proteins are shown in Fig. 7(a). The
protein level from TF gene is now bimodal in nature though the mean
level remains same as before (Fig. 2(a)). The protein level from the
functional gene is now more fluctuating about a mean value $420$
with SD = $40.05$ (Fig. 7(a) and 7(b)). The number of transitions
between inactive to active state $n_{T}$ is $130$ ($n_{F}=1612$).
The average approximate energy cost of protein synthesis is $E_{3}=A+1742H+420K$.
$E_{3}$ is lower than the $E_{1}$ and $E_{2}$. Here, the noise
in TF and functional protein levels are more than that in \noun{Strategy
I }and\noun{ Strategy II. }

\noun{Strategy IV: }Here we consider the slower transition rate constants
between the active and inactive states of TF gene compared to the
protein's degradation rate constant ($k_{aT}=0.002$, $k_{dT}=0.002$,
$k_{pT}=1.0$). The time evolution of TFs and functional proteins
are shown in Fig. 8(a). The time evolution shows that both genes remain
silent for a long period with very small active period. The protein
level from TF and functional genes are now bimodal in nature (Fig.
2(b) and 8(b)). At the steady state, the functional protein level
stays very short period above the critical level and a very long period
below the critical level (Fig. 8). The number of transitions between
the inactive and active states ($n_{T}=0,\:n_{F}=408$) of the genes
give the amount of energy cost $E_{4}=A+408H+350K$. $E_{4}$ is much
lower than that in other strategies. 

Numerical values of different quantities associated with the noise
properties in four strategies are shown in Table 1. The average cost
of energy is lowest in \noun{Strategy IV} but the fluctuations in
protein level from functional gene is too high. The protein level
falls below the critical value and stays there for longer time. That
kind of protein synthesis is not suitable in cellular processes as
observed in the case of haploinsufficiency \cite{key-131,key-135}.
The \noun{Strategy IV} is energetically suitable but functionally
unsuitable for cases when protein level has to stay above the critical
level. But, \noun{Strategy I}V may be helpful for cases when the functional
proteins are not required for longer period of time. In the \noun{Strategy
III, }the energy consumption is little bit higher but protein level
from functional gene always lies above the critical level. Therefore,
the\noun{ Strategy III} is most suitable compared to others. We found
that bimodal response of TFs with slow transition rates in\noun{ Strategy
III} is suitable to keep the protein level from functional gene above
a critical value with minimum consumption of energy. The dynamics
of TFs modulate the dynamics of functional gene states. Because of
the slower dynamics of transcription and degradation of functional
proteins and moderate transitions between ON and OFF states, the functional
proteins never come to very low level (basal level) rather always
stay above the critical level. The assumption that the protein synthesis
and degradation dynamics for functional gene is slower than the TF
gene is crucial for our result. Many studies show that the synthesis
and degradation dynamics of proteins are slower than the dynamics
of gene states \cite{key-8,key-10,key-11}. 

\begin{table}
\begin{centering}
\begin{tabular}{|c|c|c|c|c|c||c|c}
\hline 
\noun{Strategy} & $k_{aT}$ & $k_{dT}$ & TF Response & TFs FF (SD) & $n_{T}$ & $n_{F}$ & Mean FP (SD)\tabularnewline
\hline 
\hline 
I & 80 & 80 & Unimodal & 2.14 (24.2) & 47846 & 2343 & 548 (25.44)\tabularnewline
\hline 
II & 4 & 4 & Unomodal & 21.4 (76.8) & 2372 & 2206 & 513 (27.25)\tabularnewline
\hline 
III & 0.2 & 0.2 & Bimodal & 132.5 (190.8) & 130 & 1612 & 420 (40.05)\tabularnewline
\hline 
IV & 0.002 & 0.002 & Bimodal & 184.3 (225.1) & 0 & 583 & 350 (187.20)\tabularnewline
\hline 
\end{tabular}
\par\end{centering}
\caption{ Neumerical values in that table are obtained from our simulation
using GA for both the genes for four different strategies. The Hill
coefficient is set at $n=4.$ The counting is started from $t=800.$
$n_{T}$ is the number of transitions from the inactive to active
states for the TF gene. $n_{F}$ is the number of transitions from
inactive state to active states for the functional gene. The mean
functional protein (FP) level for the different Strategy is gradually
decreasing. }
\end{table}

The\noun{ }four different strategies considered here basically represent
four different probable situations of regulatory molecules or TFs
in the cell. \noun{Strategy} I (\noun{Strategy II)} represent the
situation such that the regulatory molecules are always present with
large number with a little (large) noise about a steady value. Again,
regulatory molecules for a gene may not be present with large number
continuously and throughout the time rather than they remain present
for regulation for a short period followed by a short period of absent
or low/basal value. That situation is represented by \noun{Strategy
III (F}ig. 7(a)\noun{). }It may also happen that regulatory molecules
remain absent for regulation for a longer period and become available
only for a very short period. That situation is represented by \noun{Strategy
IV.} The results show that the short duration of availability and
unavailability of regulatory molecules (i.e., large noise) for the
regulation of functional gene is suitable to keep the protein level
above the critical level with low energy consumption. The high and
low levels of TFs considered here are arbitrary. The low level can
be zero and high lvel can be a small number but greater than the Hill
coefficient $n$. Small variation in the number of regulatory molecules
gives rise to large fluctuations when the copy number of that molecules
is low. Thus, cell can produce fluctuating protein level with minimum
cost of energy by noisy low copy number of regulatory molecules.

In liver, $G6PC$ gene plays an important role in glucose homeostasis.
In the fasting condition the blood glucose level becomes low. The
proteins from $G6PC$ gene then helps to convert the stored energy
in the liver and release it into the bloodstream to raise the blood
glucose level. Whereas in fed condition, the blood glucose level is
high and then liver removes extra glucose from bloodstream to store
it again in the liver. Halpern et al. found that in fed condition,
the $G6PC$ gene expression is very infrequent with very large OFF
period and very small ON period of protein synthesis whereas in the
fasting condition, the random switching between ON and OFF period
is higher than the fed condition \cite{key-12,key-152}. In the fed
condition, the protein synthesis from the $G6PC$ are no longer essential
with higher level and cell shuts it down for longer period. In the
fasting condition bloodstream requires glucose and $G6PC$ gene do
that job by converting and transforming it from the liver. The behaviour
of $G6PC$ gene in the experiment can be compared with the behaviour
of functional gene in our simulation study. In fed condition, the
gene adopt the \noun{Strategy IV }whereas in fasting condition the
cell may adopt any strategies between I to III depending on the situation.
In the experiment, they also observed that the burst of mRNA synthesis
from the $G6PC$ gene increases and the degradation rate decreases
to raise the accumulation of mRNA in the cell. This experimental observation
clearly indicates the existence of critical or threshold level of
protein for its proper functioning. In the fasting condition, the
$G6PC$ gene is not ON or active continuously rather switching between
ON and OFF states so that protein level can fluctuates also. Since,
cell itself changes the production and degradation of mRNA synthesis
to convert the glucose from liver to bloodstream, so it is desirable
that mRNA level should not be too high rather be very close to the
threshold value. So, in fasting situation the $G6PC$ gene may follow
the\noun{ Strategy III.}

Acar et al. \cite{key-151} showed that, in a rapid fluctuating environment,
cell population's growth rate is higher for fast promoter switching
of the gene rather than the slow switching cells. In fast switching
process, cells takes more foods i.e., consume more energy and respond
faster than the slow switching process in rapidly changing environment.
The signal from environment determines the TFs level in cell through
series of biochemical events \cite{key-13,key-30}. Fluctuating environment
also gives rise to fluctuation in TF numbers. The study of Acar et
al. and Halpern et al. showed that cell can adopt any strategy depending
on its situation and environmental conditions.

\section{Conclusion}

The cell has a energy cost for the synthesis of proteins from a gene
\cite{key-140,key-141,key-142,key-143}. The cost has different values
for the different steps depending on the complexity of the steps \cite{key-10}.
It is shown by Huang et al. \cite{key-10} that the gene activation
and deactivation is costlier also. The energy input is necessary and
important to carryout each step of the GE process. It is also known
that the cellular system has evolved itself in such a way so that
it can minimize the energy cost for its activity and maximize the
outcome \cite{key-142,key-143,key-144}. The strategy of optimization
principle is followed in cellular processes to carryout its functional
activity. The cellular system can choose or adjust the reaction rates
to save energy consumption during protein synthesis and also works
reliably \cite{key-12}.

We studied a simple gene regulatory network with two genes qualitatively.
The functional gene is regulated by the proteins from the TF gene.
Different expression and noise level of proteins (different strategies)
are possible depending on the rate constants of different steps of
gene expression of the two genes. We observed the dependence of noisy
expression level of functional proteins on the fluctuation of TF proteins.
We consider four different strategies depending on low to high fluctuation
in TF proteins. In the \noun{Strategy I} (Fig. 5), we see that TF
molecules are always present with high level and with a little fluctuation
about a steady level. As a result, the functional gene has more number
of flips between active and inactive states and high mean protein
level. This behaviour is the utmost beneficial for the functioning
of the network though that requires maximum energy cost. In \noun{Strategy
II} (Fig. 6), TF molecules are always present though with higher fluctuation
about a steady level. That decreases the number of flips between the
gene states and the mean functional protein level thereby lowering
the energy cost of the cell. In \noun{Strategy III} (Fig. 7), the
TF proteins are not available always with high level to regulate the
functional gene rather they present in high level with very short
period followed by low level (or absent) with very short period also.
When TF molecules are at low level, the functional gene turns into
OFF state and as a result, the accumulated proteins degrade only.
Now, as the TF molecules move to high state, the functional gene also
turns into ON state and synthesis starts. Thus, the flips number and
mean protein level from functional gene is controlled by high randomness
of the regulatory molecules. The energy cost for protein synthesis
using this strategy is lowered than the \noun{Strategy} I and\noun{
Strategy} II. Though the functional protein level is fluctuating but
always lies well above the critical level. So, the TF molecules need
not to be present always at the higher level to maintain the functional
protein level above the threshold value. That became possible due
to the assumption of very low degradation rate constant of the proteins
from the functional gene. The protein level from functional gene does
not come down to very low level during the OFF period of TF gene and
functional gene. The \noun{Strategy IV} (Fig. 8) is not suitable to
maintain the protein level from functional gene above some threshold
value because it goes down below the threshold level and stays there
for longer time due to long time inactivity of the TF gene. But, this
is suitable for cases when long OFF periods are essential \cite{key-12}.
The scenario in \noun{Strategy} III is similar like that of the modern
day's refrigerator. The refrigerator automatically switches off its
power supply when not required, thereby reducing the energy consumption.
In gene expression and regulation processes, the transcription factors
are shared by multiple genes for their regulation and that kind of
sharing also creates noise in mRNA and protein level \cite{key-150}.
The involvement or binding of some TFs for the regulation of one gene
means their unavailability for the regulation of some others genes.
This is also a kind of 'switch-off' condition of the regulatory molecules
of the gene to whom that TFs are essential for its regulation. Thus,
cells can efficiently maintain a required protein level with fewer
number of regulatory molecules. Employing two different kinds of regulatory
molecules of opposite nature (activator and repressor) the noise and
mean level of functional molecules can also be controlled \cite{key-24,key-25,key-29}.
Therefore, our important observation is that, cellular system produces
fluctuating protein level to save energy consumption simply by employing
low copy number of regulatory molecules. Thus, by creating low copy
number of regulatory molecules in the cells the protein level and
the noise in target gene expression can be controlled efficiently.
The cell can adjust the protein level from functional gene above the
critical level by adjusting the synthesis and degradation rate constants
as well as the number or noise in regulatory molecules with minimum
energy cost.

\end{document}